\documentclass[3p,times,twocolumn]{elsarticle}

\usepackage{ecrc}

\volume{00}

\firstpage{1}

\journalname{Nuclear Physics B Proceedings Supplement}

\runauth{}

\jid{nuphbp}

\jnltitlelogo{Nuclear Physics B Proceedings Supplement}

\usepackage{amssymb}

\usepackage[figuresright]{rotating}

\def\bef{\begin{figure}}
\def\eef{\end{figure}}

\newcommand{\be}[1]{\begin{equation}\label{#1}}  
\newcommand{\beq}{\begin{equation}}
\newcommand{\eeq}{\end{equation}}
\def\ee{\end{equation}}
\newcommand{\beqn}[1]{\begin{eqnarray}\label{#1}}
\newcommand{\eeqn}{\end{eqnarray}}
\newcommand{\bd}{\begin{displaymath}}
\newcommand{\ed}{\end{displaymath}}

\newcommand{\mat}[4]{\left(\begin{array}{cc}{#1}&{#2}\\{#3}&{#4}
\end{array}\right)}

\def\lsim{\raise0.3ex\hbox{$\;<$\kern-0.75em\raise-1.1ex
\hbox{$\sim\;$}}}
\def\gsim{\raise0.3ex\hbox{$\;>$\kern-0.75em\raise-1.1ex
\hbox{$\sim\;$}}} 
\def\simlt{\mathrel{\lower2.5pt\vbox{\lineskip=0pt\baselineskip=0pt
           \hbox{$<$}\hbox{$\sim$}}}}
\def\simgt{\mathrel{\lower2.5pt\vbox{\lineskip=0pt\baselineskip=0pt
           \hbox{$>$}\hbox{$\sim$}}}}
\def\unity{{\hbox{1\kern-.8mm l}}}

\def\La{\Lambda} 

\def\lpr{l^\prime}
\def\phpr{\phi^\prime}

\def\barl{\bar{l}} 
\def\barphi{\bar{\phi}} 
\def\barlpr{\bar{l}^\prime} 
\def\barphpr{\bar{\phi}^\prime} 
\newcommand{\ov}{\overline}
\renewcommand{\to}{\rightarrow}

\begin{document}

\begin{frontmatter}



\dochead{}

\title{Shadow dark matter, sterile neutrinos and 
neutrino events at IceCube }


\author{Zurab Berezhiani}

\address{Dipartimento di Fisica, 
Universit\'a di L'Aquila  and INFN, Laboratori Nazionali del Gran Sasso, L'Aquila, Italy}

\begin{abstract}
The excess of high energy neutrinos observed by the IceCube collaboration 
  might originate from baryon number violating decays 
of heavy shadow  baryons from dark mirror sector which produce shadow neutrinos. 
These  sterile neutrino species then oscillate into ordinary neutrinos 
transferring to them specific  features of their spectrum. In particular, this 
scenario can explain the end of the spectrum above 2 PeV or so and the 
presence of the energy gap between 400 TeV and 1 PeV.  
\end{abstract}




\end{frontmatter}




Recently the IceCube Collaboration published 
the data on high-energy neutrinos  collected between 2010 and 2013, 
containing 35 candidate events in the energy range 
from 30 TeV to 2 PeV, which show  an evident excess over the expected background 
of the events  with $E > 60-100$~TeV or so \cite{IC}. 
On the other hand, no events were observed 
in the gap between 400 TeV and 1 PeV while   
three most energetic shower events emerged 
at the end of the spectrum with energies between 1-2 PeV  
 where the atmospheric background is practically vanishing.  
The spectrum is apparently cut off at energies larger than about 2 PeV. 
The gap in the energy spectrum is difficult to explain in known models 
of high-energy neutrinos of astrophysical origin. 

Here we present  a  model \cite{BBDG} that may explain such a spectrum. 
It is based on the idea that dark matter of the universe 
emerges from a parallel gauge sector,  
with particles and interactions sharing many similarities with ordinary particle sector.  
Such a shadow sector would contain particles like quarks which form 
composite baryons, as well as leptons and neutrinos 
which are all sterile for ordinary gauge interactions.  
Particularly interesting example is represented by  
so-called mirror world \cite{Berezhiani:2003xm}, 
which has the particle and interaction content exactly identical to that of ordinary sector, 
with the same gauge and Yukawa coupling constants. 

Taking into consideration also attractive possibilities for  physics beyond the Standard Model 
related to supersymmetric (SUSY) grand unified theory  (GUT),   
one can consider that    
at higher energies our physics is presented by SUSY GUT,  
 e.g. $SU(5)$  or $SU(6)$ which breaks down to the Standard Model 
 $SU(3)\times SU(2) \times U(1)$ at the scale $M_G \simeq 2\times 10^{16}$~GeV. 
Supersymmetry breaking at $M_{\rm SB} \sim 1$~TeV triggers the electroweak symmetry 
breaking and the Higgs field gets the vacuum expectation value (VEV)  $v = 174$~GeV. 
In this view, we assume  that   at higher energies also mirror sectors is presented 
by the identical SUSY GUTs, $SU(5)'$ or $SU(6)'$, 
which  breaks down to its standard subgroup $SU(3)'\times SU(2)' \times U(1)'$ 
at the same scale $M_G \simeq 2\times 10^{16}$~GeV. 
However, following refs. \cite{Berezhiani:1995, Akhmedov:1992hh}, 
we  assume that  the symmetry between two sectors is broken later 
so that the electroweak symmetry breaking scale $v'$ in mirror sector is 
much larger than ordinary electroweak scale. 
Namely, if $v'\sim 10^{11}$~GeV,  the lightest shadow baryons 
have masses order few PeV, and they decay  
due to baryon violating GUT gauge bosons, 
with decay time comparable to the age of the Universe, 
producing  energetic shadow neutrinos which then oscillate into active neutrinos 
(with oscillation probablities $\sim 10^{-9}$ or so) 
transferring their spectrum to the latter.\footnote{For other type of decaying dark matter 
model see e.g.  Ref. \cite{Esmaili:2013gha}. } It is worth to note that the decaying dark 
matter model, with a fraction of dark matter of about 10 per cent decaying before of present epoch 
could reconcile the Planck collaboration results on the CMB measurements with 
low redshift astrophysical measurements \cite{Berezhiani:2015yta}.

In other words, we consider  a supersymmetric grand unification theory $SU(5) \times SU(5)'$
or $SU(6) \times SU(6)'$.  
As discussed in ref. \cite{BBDG}, our proposal can be more nicely realized in 
SUSY $SU(6)$ theory \cite{Berezhiani:1989bd} 
which gives natural solution to the so called hierarchy and doublet-triplet splitting problems  
via the Goldstone boson mechanism, 
relating the electroweak symmetry breaking scale to the supersymmetry 
breaking scale $M_{\rm SB}$, and in addition naturally explains the fermion mass spectrum.  
In both sectors the GUT symmetries are broken at the scale 
$M_G \simeq 2\times 10^{16}$ GeV. 
Below this scale our sector is represented by the minimal SUSY Standard Model (MSSM) 
$SU(3)\times SU(2) \times U(1)$ with chiral superfields of 
quarks $q_i = (u,d)_i$, $u^c_i, d^c_i$ and leptons $l_i = (\nu,e)_i$, 
$e^c_i$  ($i=1,2,3$ is family index)  and two Higgs superfields $h$ and $\bar h$,  
described by the Yukawa superpotential 
$W = Y^e_{ij} l_i e^c_j h  + Y^d_{ij} q_i d^c h  + Y^u_{ij} q_i u^c_j \bar h$.  
Supersymmetry is then broken at the scale $M_{\rm SB} \sim 1$~TeV inducing also 
the electroweak symmetry breaking. 

As for parallel mirror sector, below the scale $M_G \simeq 2\times 10^{16}$ GeV 
we have supersymmetric $SU(3)'\times SU(2)' \times U(1)'$ theory with 
the similar particle content, quarks $q'_i = (u',d')_i$, $u^{c\prime}_i, d^{c\prime}_i$,  
leptons $l'_i = (\nu',e')_i$, 
$e^{c\prime}_i$,  and two Higgs superfields $h'$ and $\bar{h}'$. 
At the scale $M_G$ mirror gauge coupling constants  $g'_{3,2,1}$ 
are equal to the ordinary gauge constants $g_{3,2,1}$, 
and coupling constants in the Yukawa superpotential 
$W'= Y^e_{ij} l'_i e^{c\prime}_j h' + 
Y^d_{ij} q'_i d^{c\prime} h' + Y^u_{ij} q'_i u^{c\prime}_j \bar{h}'   $ 
have exactly the same pattern as in $W$.
We assume, however, that in mirror sector supersymmetry and electroweak symmetry 
are both broken at the scale of about $10^{11}$ GeV.  

\begin{figure}[b]
 \centering
\includegraphics[scale = .68]{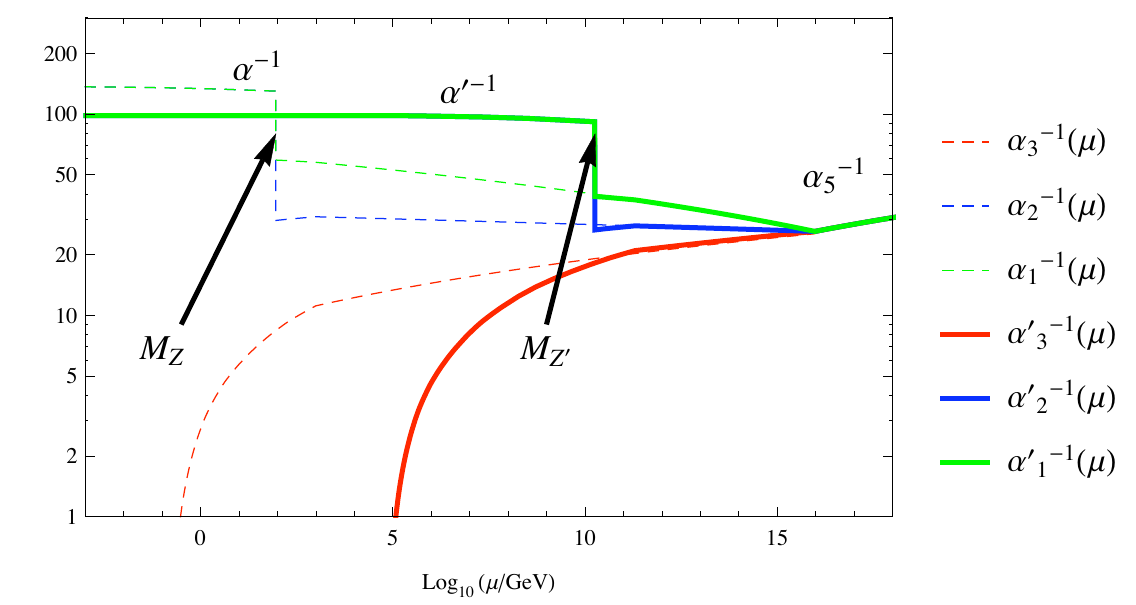}    
\caption{RG running of gauge couplings 
below the GUT scale in ordinary and shadow sectors, 
$\alpha_i = g_i^2/4\pi$ and $\alpha'_i =  g_i^{\prime 2}/4\pi$. }
\label{running}
\end{figure}

Hence, below GUT scale, the gauge coupling constants  $g_{3,2,1}$ and  $g'_{3,2,1}$ 
evolve down in energies in both sectors in the same way up to scales 
of about $10^{11}$ GeV where the supersymmetry is spontaneously 
broken in mirror sector (see Fig. \ref{running}). 
However, below this scale ordinary sector still remains  supersymmetric and 
constants evolve down by the renormalization group (RG) equations as in the MSSM, 
down to scale  $M_{\rm SB} \sim 1$~TeV where supersymmetry is effectively broken.  
After that the Higgses $h$ and $\bar h$ 
are not protected anymore by the supersymmetry and they 
get VEVs $v_1 = v\cos\beta$, $v_2 = v\sin\beta$, $v=174$~GeV, 
which induce the electroweak symmetry breaking and generate the fermion masses. 
The masses of lightest fermions, $m_e = 0.5$~MeV, $m_u\simeq 3$~MeV and 
$m_d \simeq 5$~MeV respectively for electron, up-quark and down quark, 
are related to smallness of the Yukawa constants of the first generation. 
At the QCD scale $\Lambda\simeq 200$~MeV gauge interactions of $SU(3)$ 
become strong and confine the quarks into baryons, with lightest ones 
being proton and neutron with masses of about 1 GeV and spin 1/2.

Let us consider now parallel mirror sector 
where supersymmetry is broken  at the scale 
$M'_{\rm SB} \sim 10^{11}$~GeV, 
supposedly due to non-zero $F$ or $D$ terms of some auxiliary fields. 
Hence shadow scalars, including squarks and sleptons as well as the gauginos and 
the Higgs doublets $h'$ and $\bar{h}'$ all acquire soft masses order $M_{\rm SB}$.
Respectively shadow Higgses can get VEVs $v'_1 = v'\cos\beta'$ 
and $v'_2 = v'\sin\beta'$, which break mirror electroweak symmetry at the scale 
$v'\leq M'_{\rm SB}$.\footnote{Interestingly, if 
supersymmetry breaking is transferred to our sector via gravity or 
other Planck scale mediators, this would nicely explain ordinary soft masses order 
$M_{\rm SB} \sim M^{\prime 2}_{\rm SB}/M_{Pl}\sim 1$~TeV. }
Therefore, the masses of shadow fermions  are rescaled, modulo 
renormalization factors order 1, by a factor $\zeta = v'/v$ with respect to ordinary fermion masses. 
Namely, taking $v'/v = 10^{9}$ and assuming $\tan\beta'=\tan\beta$, 
by the RG running of gauge and Yukawa constants from the GUT scale down in energies, 
one obtains for the  masses of lightest mirror fermions 
$M_E \simeq 0.4$ PeV, $M_D \simeq 1.1$ PeV and $M_U \simeq 1.9$ PeV, 
where capital letters $E,D,U$ denote respectively the 
shadow electron $e'$, down quark $d'$ and up quark $u'$.   
(Notice, that in mirror sector up-quark  becomes heavier that in down-quark due 
to the difference in the RG running of the Yukawa constants \cite{BBDG}.) 
One the other hand, the RG evolution of the $SU(3)'$ gauge constant $g'_3$ 
shows that the shadow QCD scale becomes $\Lambda' \sim 100$ TeV  
(c.f. $\Lambda \simeq 200$~MeV in ordinary QCD), as it is shown on Fig. \ref{running}.  
Therefore, $ M_U, M_D \gg \Lambda'$ 
and the shadow QCD looks like a rescaled version of our QCD but without light quarks, 
containing only the heavy quarks like $c$ and $b$.  In fact,   
$M_{U}$ and $M_D$ are larger than $\Lambda'$ by about the same factor  
as the ordinary beauty and charm quark masses, $m_{b}$ and $m_c$, 
are larger with respect to  ordinary QCD scale $\Lambda$.

As far as in shadow sector 
up quark $U$ is heavier than down quark $D$, 
 the lightest shadow baryon should be 
shadow $\Delta^-$ baryon of spin $3/2$, consisting of three down quarks $D$  
and having mass $M_\Delta \approx 3 M_D = 3.3$~ PeV.   
All states, containing up quark $U$, will be unstable against weak decays, 
$U \to D \bar{E} \nu'$. 
As for mesons, 
the lightest pseudoscalar is shadow neutral pion $\pi^0$ consisting of  $D \bar{D}$, 
with mass $M_0 \approx 2 M_D = 2.2$~PeV, while the lightest vector meson 
$\rho^0(D \bar{D})$  is slightly heavier than $\pi^0$. 
(Another neutral pion consisting of $U \bar{U}$ becomes much heavier, with mass 
of about 3.8 PeV.)
Charged Pion $\pi^\pm$  as well as $\rho^\pm$-meson  consisting of $D\bar U$
will have mass $M \simeq  M_U + M_D = 3$~PeV,  with $\rho^-$ a bit heavier than $\pi^-$. 
All pseudoscalar  and vector mesons have excited states with mass gap order 
$\Lambda'$ between the levels, just like $c\bar c$ or $b\bar b$ states in our QCD. 

Now we come to the role of baryon violation and proton decay which is fundamental 
prediction of the GUTs. The heavy gauge bosons of $SU(5)$ 
with baryon violating couplings between quarks and leptons induce the decay of 
the lightest ordinary baryons (proton, or neutron bound in nuclei),   
with lifetime $\tau_p \sim M_G^4(\alpha_G^2 m_p^5)^{-1}\sim 10^{31}$~Gyr or so, 
 where  $\alpha_G$ is gauge coupling constant at the  GUT scale 
 $M_G \sim 2\times 10^{16}$ GeV \cite{Nath:2006ut}.
In the shadow sector, the similar couplings of GUT gauge bosons should 
destabilize the shadow $\Delta$ baryon. 
However, taking into account that the latter 
is much heavier than the ordinary proton,  $M_\Delta/ m_p \sim 10^6$, 
its lifetime must be about 30 orders of magnitude smaller than the proton lifetime. 
Hence we get $\tau_\Delta \sim M_G^4(\alpha_5^2 m_\Delta^5)^{-1}\sim 100$~Gyr or so, 
comparable to the age of the Universe $t_U = 14$ Gyr.

The principal decay mode of $\Delta$ baryon is in vector mesons, 
$\Delta^- \to \rho^-_{a} + \bar{\nu}'_x$, where  generically
$\nu'_x$ is  a superposition of shadow neutrino flavor eigenstates $\nu'_{e,\mu,\tau}$. 
%
Each decay produces monoenergetic neutrinos $\nu'_x$ with energies 
$E_i = \frac12 M_\Delta (1 - M_i^2/M_\Delta^2)$, where $M_0$ is the mass of 
$\rho^-$ meson and $M_{1,2,...}$ are the masses of its excitations. 
In Fig. 2(a) the spectrum of 
neutrinos produced by decay of galactic dark matter is shown by sharp peaks 
(solid blue) for $M_0, M_1, M_2$ respectively being  $3.0, 3.1$ and 3.2~PeV. 
Due to close degeneracy between the masses of $\Delta$-baryon and 
$\rho^-_a$ mesons, the neutrino energies are $E_i \ll 1$ PeV while   
the most of initial energy $=M_\Delta$ 
is taken away by  vector mesons $\rho^-_i$.

\begin{figure}[t]
 \centering
{\includegraphics[scale = 0.42]{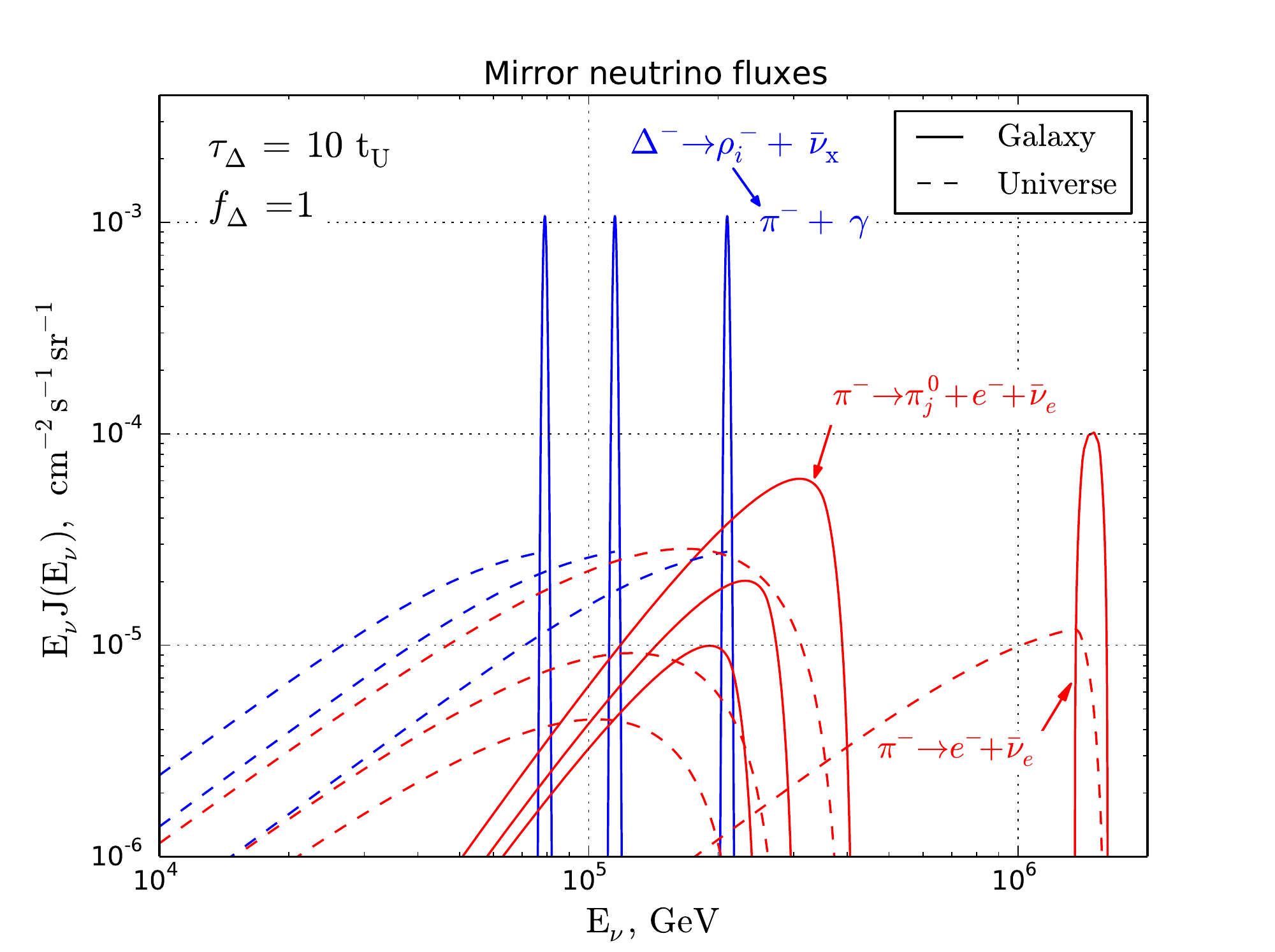}}
 \hspace{5mm}
{\includegraphics[scale = 0.42]{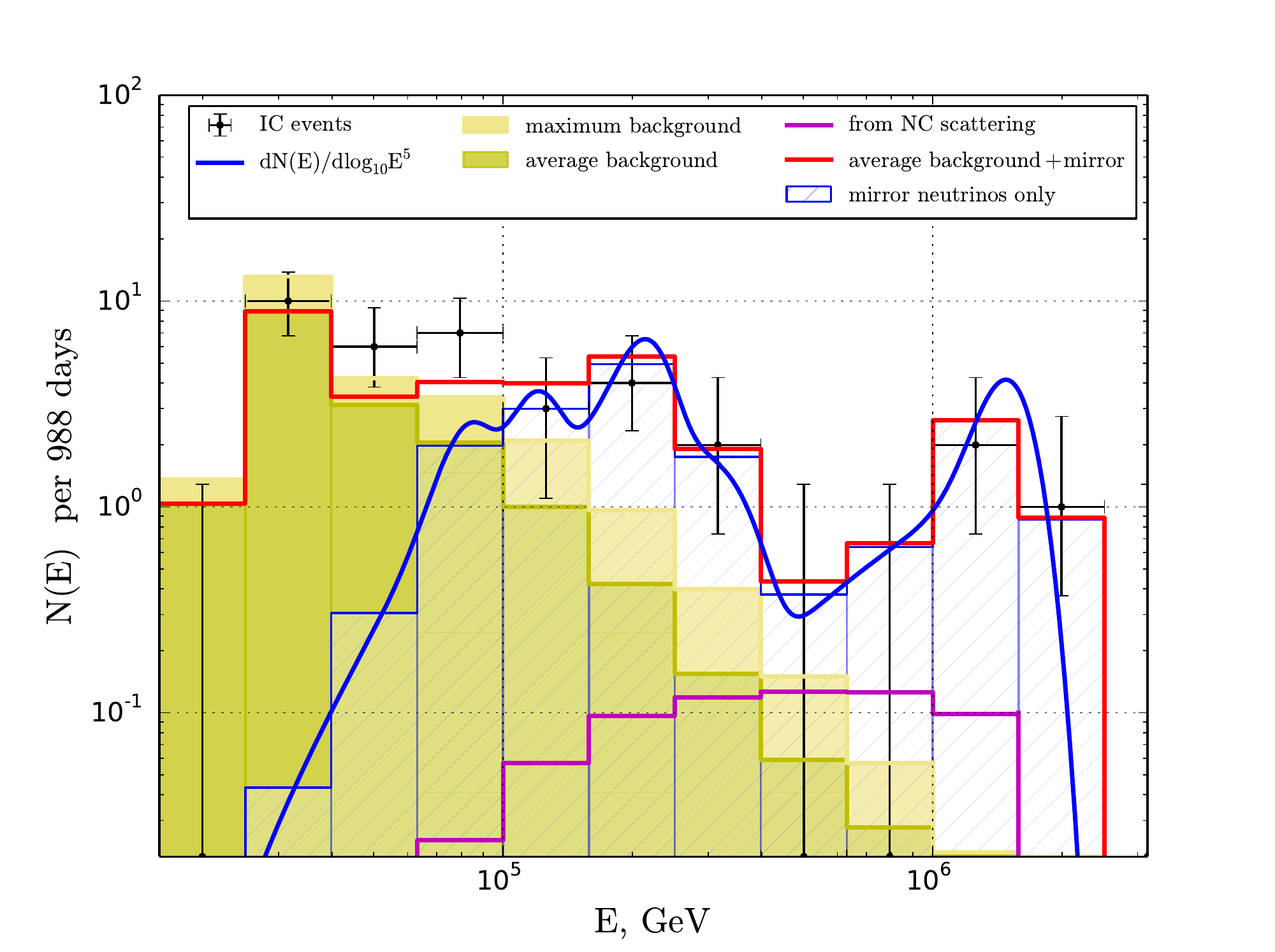}}
 \caption{(a) Shadow neutrino spectrum produced by $\Delta$-barion decay  
 and subsequent prompt decays of shadow pions 
  (b) Spectrum of VHE neutrino events at IceCube as predicted in our model. Magenta line 
shows contribution of the neutral current interaction and its integral corresponds to 1 event 
in the energy interval 100 TeV -- 2 PeV. }
\label{spectrum} 
\end{figure}

However, vector mesons readily decay into mirror pion and photon, 
$\rho^-_i \to  \pi^- + \gamma'$, and subsequent decay of the pion produces the neutrino once 
again (solid red curves in Fig. 2(a)). 
Shadow $\pi^-$ has two decay modes,  two body $\pi^{-}  \to e' \bar{\nu}'_e$ and 
three body $\pi^{-}  \to \pi^0_j e' \bar{\nu}'_e$, where $\pi^0_j$, $j=0,1,...$ are the basic 
shadow pion and its excited states.  
Interestingly, the 2-body and 3-body branching ratios 
are comparable. 
This fact is intimately related to the value $\Lambda' \sim 100$~TeV \cite{BBDG}. 
Two body decay produces neutrinos with a narrow energy spectrum concentrated 
around $M_\Delta/2 \simeq 1.6$ PeV, while the three body decay, due to smaller phase space, 
produces less energetic neutrinos with a wide spectrum extending up to the value 
$E_{\rm max} = M_- - M_0 - M_E \simeq M_U - M_D - M_E \simeq 0.4$ PeV.

Fig. 2(a) shows the final spectrum of shadow neutrinos $\nu'_x$ and $\nu'_e$, 
including the neutrinos produced by the decay of 
dark matter in the galactic halo, and extragalactic neutrinos produced 
by the decay of cosmological dark matter at large redshifts. 
For definiteness, in Fig. 2(a) 
we assume that dark matter entirely consists of shadow baryons $\Delta$, 
i.e. the fraction $f_\Delta =1$, 
and take the decay time $\tau_\Delta$ as 10 times the Universe age,  
$\tau_\Delta = 10 t_U$. 
The fraction of extragalactic neutrinos strongly depends on the decay time $\tau_\Delta$, 
 it increases with $\tau_\Delta$ decreasing. In correspondence, the energy gap becomes 
 less pronounced and at $\tau_\Delta < 0.1 t_U$ or so it practically disappears since 
 cosmological contribution from  high redshifts become dominant. In this case 
dark matter cannot be entirely from  mirror sector, the later can constitute 
only a smaller fraction and other type of stable dark matter should be also invoked.\footnote{  
It is worth to note that the decaying dark 
matter model, with a fraction of dark matter of about 10 per cent decaying before of present epoch 
could reconcile the Planck collaboration results on the CMB measurements with 
low redshift astrophysical measurements \cite{Berezhiani:2015yta}. }

The shadow neutrinos may have mixing with ordinary ones \cite{Berezhiani:1995}. 
We assume that operators mixing $\nu_{e,\mu,\tau}$ and $\nu'_{e,\mu,\tau}$ states 
respect a conservation of a combined  lepton number $\bar L = L - L'$ \cite{Berezhiani:2005hv}, 
and  that all operators for neutrino masses are suppressed by the Planck scale 
$M_{Pl}$ \cite{Akhmedov:1992hh}. Hence, for neutrino masses and mixing we 
consider the following  operators (family indices are suppressed):
\be{nus}  
\frac{A\chi}{M^2_{Pl}} l l h h  +  \frac{A\ov{\chi}}{M^2_{Pl}} l' l' h' h'  
+ \frac{D}{M_{Pl}} l l'  h h'  
\ee 
where $\chi$ and $\bar{\chi}$ are gauge singlet chiral superfields  
with lepton numbers $\bar L = -2$ and 2 respectively, with VEVs 
$\langle \chi\rangle = \langle \bar \chi\rangle = \mu$ breaking 
the lepton number. Alternatively,  $\chi$ and $\ov\chi$  can be promoted 
as flavon sextet and anto-sextet fields of the flavor symmetry $SU(3)_H$ 
\cite{Berezhiani:1983hm} between three families assuming that it is a common 
gauge symmetry between two sectors \cite{Berezhiani:1996ii}. 
This would give an certain guideline for obtaining predictive patterns for active and 
sterile neutrino masses and mixing. 

Here the first  two terms give the Majorana masses respectively to ordinary neutrinos 
$\nu_{e,\mu,\tau}$ and their shadow (sterile) partners $\nu'_{e,\mu,\tau}$
while third term induces the mixing (Dirac) terms between active and shadow neutrinos. 
 Then total  $6\times 6$ mass matrix of  
 $\nu$ and $\nu'$  states  reads \cite{Bento:2001rc}:
 \begin{equation} \label{numass} 
M_\nu = \mat{m_\nu}{m_{\nu\nu^\prime}}{m_{\nu\nu^\prime}^T}
{m_{\nu^\prime}} = 
\frac{v^2}{M} \mat{A\lambda }{D\zeta}{D^T \zeta }
{A \lambda \zeta^{2}}  
\end{equation}
where  $\zeta = v'/v$  and $\lambda = \mu/M_{Pl}$. Therefore, taking e.g. 
$\lambda \sim 10^{-5}$ and constants $A,D\sim 10^{-2}$, 
we see that shadow neutrinos acquire masses $\sim 10$~keV 
and active sterile mixing angles are $\sim  10^{-4}$.

%

Fig. 2(b) shows how the spectrum of shadow neutrinos shown on Fig. 2(a) 
and transferred to ordinary neutrinos via active-sterile mixings 
will be seen by the IceCube. 
Here the effective areas for the neutrino detection by IceCube \cite{IC} 
and characteristic  error bars in estimation 
 of neutrino energies (of about 13$\%$) are taken into account. 
The obtained spectrum of events indeed look very much like the spectrum 
observed by the IceCube \cite{IC}.  
 In addition, this model can has specific 
predictions for the flavor content of these high energy neutrinos 
in different parts of the spectrum \cite{BBG}.  

The validity of our model can be tested with  increasing statistics 
by the IceCube collaboration.

\vspace{3mm} 
\noindent
{\bf Acknowledgements} 
\vspace{2mm} 

I thank R. Biondi, G. Di Panfilo and A. Gazizov for collaboration.  
This work was supported in part by MIUR research grant "Theoretical Astroparticle
Physics" PRIN No. 2012CPPYP7 
and in part by Rustaveli National Science Foundation grant No. DI/8/6-100/12. 
I would like to thank for the hospitality 
organizers of Int. Workshop NOW 2014, Conca Specchiula, Italy, 7-14 Sept. 2014, 
where this work was reported.

\nocite{*}
\bibliographystyle{elsarticle-num}
\bibliography{martin}



\end{document}